\documentclass[oneside]{article}
\usepackage{graphicx}
\usepackage{bbold}
\usepackage{amsfonts}
\usepackage[parfill]{parskip}
\usepackage{amsmath}
\graphicspath{{images/}}
\title{Modified Cahn model for surface tension with the presence of a contact line applied to regular solutions.}

\author{Lionel Hirschberg and Avraham Hirschberg}

\begin{document}
\maketitle

\begin{abstract}
We derive the interfacial tension of a semi infinite two phase binary mixture of partially miscible fluids in contact with a planar surface. In our model we take a surface free energy term into account omitted by Cahn [J. W. Cahn, J. Chemical Physics, 66:3667-3672,1977]. Our analysis applied to the case of regular solutions shows that this extra surface free energy term cannot a priori be neglected.
\end{abstract}
\tableofcontents

\section{Introduction}
In his paper on critical point wetting Cahn \cite {cahn_wetting} considers the free energy of a semi infinite two phase binary mixture of partially miscible fluids in contact with a planar surface at $x=0$. He assumes that the surface is sharp on an atomic scale and that the interactions between the fluid and the substrate are sufficiently short ranged such that the contribution to the free energy can be modeled by a free energy per unit surface $f_w(c_s)$. Where $c_s$ is the composition of a fluid particle adjacent to the wall ($x=0$) n.b. $c$ is the molar concentration of one of the components in the binary mixture.

Cahn \cite {cahn_wetting} uses the following expression for the wall excess free energy per unit area:

\begin{equation}\label{eq:Cahn77}
\Delta F=f_w(c_s)+\int_0^\infty \left[\Delta f +\kappa \left(\frac{d c}{dx}\right)^2\right]dx.
\end{equation}

Which essentially is the expression derived in his earlier paper\cite{Cahn_Hilliard} for the surface tension at the interface between a planar interface between two phases of a mixture of partially miscible fluids: 

\begin{equation}\label{eq:Cahn58}
\gamma=\int_{-\infty}^\infty \left[\Delta f +\kappa \left(\frac{d c}{dx}\right)^2\right]dx.
\end{equation}


In equation (\ref{eq:Cahn77}) the surface interaction potential $f_w$ replaces the integral from $x=-\infty$ to $x=0$ in equation (\ref{eq:Cahn58}).

However the expression for the surface tension in Cahn-Hilliard \cite{Cahn_Hilliard} is derived for cases were wall surface effects are negligible. Applying this expression for the cases where surface effects play an essential role is not a priori justified as we explained in our manuscript submitted to the Journal of fluid Mechanics \cite{JFMLioMico}. 

Here we derive a model for the excess free energy $\Delta F$ of a semi infinite two phase binary mixture of partially miscible fluids in contact with a planar surface that takes into account the surface term neglected by Cahn \cite{cahn_wetting}. We do this for the particular case of a regular solution model as described in the paper of Cahn and Hilliard \cite{Cahn_Hilliard}. This will allow us to compare the term neglected with other terms in the model giving us some insight into whether or not this term is justifiably neglected by  Cahn \cite{cahn_wetting}.

\section{Theory}
We define the excess free energy $\Delta F$ per unit area as the free energy per unit area of a semi infinite two phase binary mixture of partially miscible fluids in contact with a planar surface $F$ minus that of a semi infinite uniform reference system $F_0$. Hence this is the energy needed to create the fluid in contact with the wall out of a uniform reference fluid of concentration $c_0$, the molar fraction far away from the wall. The reference state is one for which the fluid has no interaction with the wall. Thus we have 

\begin{equation}
\Delta F=F-F_0,
\end{equation}

where F is

\begin{equation}\label{eq:F}
F=f_w(c_s)+\int_0^\infty f dx.
\end{equation}

Here $f$ is the free energy per unit volume which below the critical point can be expanded in a Taylor series around the uniform state which for an isotropic fluid yields \cite{Cahn_Hilliard,Lionel}

\begin{eqnarray}
f&\simeq& f_0(c)+\left(\frac{\partial f}{\partial d^2 c / d x^2}\right)_0\left(\frac{d^2 c}{d x^2}\right)+\frac{1}{2}\left(\frac{\partial^2 f}{\partial (d c / d x)^2}\right)_0\left(\frac{d c}{d x}\right)^2\nonumber\\
&=&f_0(c)+\kappa_1\left(\frac{d^2 c}{d x^2}\right)+\kappa_2\left(\frac{d c}{d x}\right)^2\label{eq:Taylor}.
\end{eqnarray}

Where $\kappa_1\equiv\left(\partial f/\partial d^2 c / d x^2\right)_0$ and $\kappa_2\equiv\left(\frac{\partial^2 f}{\partial (d c / d x)^2}\right)_0/2$, the subscript $0$ specifies functions of $c$ (viz. not of $d^2 c / d x^2$ and $\left(d c/d x\right)^2$). Substituting equation (\ref{eq:Taylor}) into equation (\ref{eq:F}) yields

\begin{equation}\label{eq:FT}
F=f_w(c_s)+\int_0^\infty \left[f_0(c)+\kappa_1\left(\frac{d^2 c}{d x^2}\right)+\kappa_2\left(\frac{d c}{d x}\right)^2\right] dx.
\end{equation}

Performing partial integration on the second term in the integrant and using 

\begin{eqnarray}
\lim_{x\rightarrow\infty}\frac{d c}{d x}&\rightarrow&0\\
\lim_{x\rightarrow0}\frac{d c}{d x}&\rightarrow&\left(\frac{d c}{d x}\right)_{c_s}
\end{eqnarray}

yields

\begin{eqnarray}
F&=&f_w(c_s)-\kappa_1\left(\frac{d c}{d x}\right)_{c_s}+\int_0^\infty \left\{f_0(c)+\left[-\frac{d \kappa_1}{d c}+\kappa_2\right]\left(\frac{d c}{d x}\right)^2\right\} dx\nonumber\\
&=&f_w(c_s)-\kappa_1\left(\frac{d c}{d x}\right)_{c_s}+\int_0^\infty \left[f_0(c)+\kappa\left(\frac{d c}{d x}\right)^2\right].
\end{eqnarray}
Where $\kappa\equiv\kappa_2-d\kappa_1/dc$. For the particular case considered i.e. the regular solution model of Cahn and Hilliard \cite{Cahn_Hilliard} we have $\kappa_1=-\kappa c$ where $\kappa$ is a positive constant and $\kappa_2=0$.
The free energy of the reference state is

\begin{equation}\label{eq:F0}
F_0=\int_0^\infty f_0(c_0) dx,
\end{equation}

where $c_0\equiv\lim_{x\rightarrow\infty} c $. Thus defining $\Delta f\equiv f_0(c)-f_0(c_0)$ the excess free energy $\Delta F$ can be written as:

\begin{equation}
\Delta F=f_w(c_s)-\kappa_1\left(\frac{d c}{d x}\right)_{c_s}+\int_0^\infty \left[\Delta f+\kappa\left(\frac{d c}{d x}\right)^2\right].
\end{equation}

with $\kappa_1=-\kappa c$. Taking the variation\footnote{We use  \begin{eqnarray*} \kappa\delta\int_0^\infty \left(\frac{d c}{d x}\right)^2 dx&=&2\kappa\int_0^\infty\frac{d c}{dx}\delta \frac{d c}{dx} dx=2\kappa\int_0^\infty\frac{d c}{dx}\frac{d }{dx}\delta c dx\\
&=&-2\kappa\left(\frac{d c}{d x}\right)_{c_s}\delta c_s-2\kappa\int_0^\infty\frac{d^2 c}{d x^2}\delta c dx.
\end{eqnarray*}} of $\Delta F$ we have

\begin{eqnarray}
\delta \Delta F&=&\left\{\frac{d f_w}{d c_s}-\frac{d}{d c_s}\left[\kappa_1\left(\frac{d c}{d x}\right)_{c_s}\right]-2\kappa\left(\frac{d c}{d x}\right)_{c_s}\right\}\delta c_s\nonumber\\
&+&\int_0^\infty \left[\frac{d \Delta f}{d c}-2\kappa\frac{d^2 c}{d x^2}\right]\delta c dx\label{eq:deltaF}.
\end{eqnarray}

In equilibrium we have $\delta \Delta F=0$ where surface term of the variation and the bulk term should vanish independently \cite{CourantH}. Setting $\delta c_s=0$ and $\delta c\neq0$ we find the equilibrium condition for the bulk

\begin{equation}
\frac{d \Delta f}{d c}=2\kappa\frac{d^2 c}{d x^2}.
\end{equation}

We can integrate this ordinary differential equation using the boundary condition $\Delta f=0$ and $d c / d x=0$ in the limit $x\rightarrow\infty$ to find:

\begin{equation}
\Delta f=\kappa\left(\frac{d c}{d x}\right)^2.
\end{equation}

Which can be rewritten to find 

\begin{equation}\label{eq:Bulk}
\frac{d c}{d x}=-\sqrt{\frac{\Delta f}{\kappa}}
\end{equation}

where we have chosen the negative root because we take $c_s>c>c_0$ in this model\footnote{Equation (\ref{eq:Bulk}) derived here is the same as equation (7) derived by Cahn  \cite{cahn_wetting} for the bulk.} (the wall is partially wetting) n.b. $c\in\Omega\equiv\{c\,|\,0\leq c\leq 1\}$.

The surface tension $\gamma$ is the minimum in excess free energy of the surface per unit area  \cite{cahn_wetting} viz.

\begin{eqnarray}
\gamma&=&f_w(c_s)-\kappa_1\left(\frac{d c}{d x}\right)_{c_s}+\int_0^\infty 2\kappa\left(\frac{d c}{d x}\right)^2 dx\nonumber\\
&=&f_w(c_s)-\kappa_1\left(\frac{d c}{d x}\right)_{c_s}+\int_{c_s}^{c_0} 2\kappa\left(\frac{d c}{d x}\right) dc\nonumber\\
&=&f_w(c_s)-\kappa_1\left(\frac{d c}{d x}\right)_{c_s}+\int_{c_0}^{c_s} 2\sqrt{\kappa\Delta f} dc
\end{eqnarray}

Which except for the extra surface term $-\kappa_1\left(d c/d x\right)_{c_s}$ corresponds to equation (10a) of Cahn \cite{cahn_wetting}. Using $f_w(c_s)=f_w(c_0)+\int_{c_0}^{c_s}\left(d f_w / d c_s\right) dc$ we can rewrite this equation as follows

\begin{equation}\label{eq:gamma}
\gamma=f_w(c_0)-\kappa_1\left(\frac{d c}{d x}\right)_{c_s} +\int_{c_0}^{c_s}\left[\frac{d f_w}{d c_s}+ 2\sqrt{\kappa\Delta f} \right]dc.
\end{equation}

This equation corresponds to equation (10b) of Cahn \cite{cahn_wetting} with an added surface free energy term $-\kappa_1\left(d c/d x\right)_{c_s}$.

Setting $\delta c=0$ and $\delta c_s\neq0$ in equation (\ref{eq:deltaF}) yields

\begin{equation}\label{eq:NB1}
\frac{d}{d c_s}\left(\kappa_1\left(\frac{d c}{d x}\right)_{c_s}\right)+2\kappa\left(\frac{d c}{d x}\right)_{c_s}=\frac{d f_w}{d c_s}
\end{equation}

the so called natural boundary condition at the wall

We assume $f'_w<0$ i.e. the surface is partially wetting. For simplicity we assume $f_w'=\text{constant}$. The regular solution model of Cahn-Hilliard \cite{Cahn_Hilliard} results in

\begin{eqnarray*}
\kappa_1&=&-\kappa c\,\,\forall c\in\Omega\\
\kappa_2&=&0
\end{eqnarray*}

Substituting this into equation (\ref{eq:NB1}) and defining $\varsigma\equiv\left(d c / d x\right)_{c_s}$ as well as $f_w'\equiv d f_w / d c_s$, we have

\begin{equation}\label{eq:NB2}
\frac{d \varsigma}{ d c_s}-\frac{1}{c_s}\varsigma=-\frac{f_w'}{\kappa c_s}.
\end{equation}

Which is a first order ordinary differential equation with solution $\varsigma=\varsigma_h+\varsigma_p$ i.e. the solution is composed of a homogeneous and a particular solution. The homogeneous solution can be found by omitting the forcing term $f_w'/(\kappa c_s)$ in equation (\ref{eq:NB2}) viz. solving 

\begin{equation}\label{eq:NB3}
\frac{d \varsigma_h}{ d c_s}-\frac{1}{c_s}\varsigma_h=0
\end{equation}

yields the homogeneous solution 

\begin{equation}
\varsigma_h=Dc_s
\end{equation}

where $D$ is an integration constant.

The particular solution is

\begin{equation}
\varsigma_p=\frac{f_w'}{\kappa}
\end{equation}

this can easily be verified by substitution into equation (\ref{eq:NB3}). 

We now have

\begin{equation}
\varsigma=Dc_s+\frac{f_w'}{\kappa}.
\end{equation}

We assume that for $c_s=1$ \cite{Lionel}

\begin{equation}
\left(\frac{d c}{d x}\right)_{c_s=1}=0
\end{equation}

thus we find

\begin{equation}\label{eq:fwprimesolution}
\varsigma \equiv\left(\frac{d c}{d x}\right)_{c_s}=\frac{f_w'}{\kappa}\left(1-c_s\right).
\end{equation}

We will use this to investigate the importance of the surface term $-\kappa_1\left(d c/d x\right)_{c_s}$  in equation (\ref{eq:gamma}). That is we have

\begin{eqnarray}
\gamma&=&f_w(c_0)-\kappa_1\left(\frac{d c}{d x}\right)_{c_s} +\int_{c_0}^{c_s}\left[\frac{d f_w}{d c_s}+ 2\sqrt{\kappa\Delta f} \right]dc\nonumber\\
&=&f_w(c_0)+\kappa c_s\left(\frac{d c}{d x}\right)_{c_s}+\left(c_s-c_0\right)\frac{d f_w}{d c_s} +\int_{c_0}^{c_s} 2\sqrt{\kappa\Delta f} dc\label{eq:gammacomp}.
\end{eqnarray}

where we have used that $d f_w / d c_s$ is a constant. Substituting equation (\ref{eq:fwprimesolution}) into equation (\ref{eq:gammacomp}) yields

\begin{equation}\label{eq:gammacomp1}
\gamma=f_w(c_0)+c_s\left(1-c_s\right)\frac{d f_w}{d c_s}+\left(c_s-c_0\right)\frac{d f_w}{d c_s} +\int_{c_0}^{c_s} 2\sqrt{\kappa\Delta f} dc.
\end{equation}

The second term on the right hand side of equation (\ref{eq:gammacomp1}) is a extra surface free energy term omitted by Cahn \cite{cahn_wetting}. 

\section{Discussion}

We are know going to compare the term neglect by Cahn \cite{cahn_wetting} to the third term on the right hand side of equation (\ref{eq:gammacomp1}), which is a bulk term also present in Cahn's expression \cite{cahn_wetting}, taking the fraction of the extra surface term with respect to the bulk term yields

\begin{equation}
\Lambda\equiv\frac{c_s(1-c_s)}{(c_s-c_0)}=\frac{1-c_s}{1-\frac{c_0}{c_s}}.
\end{equation}

%
Which in general will not be small for $c_s\in(0,1)$ and $c_0<c_s$.

We conclude that one cannot make the a priori assumption that the results obtained by Cahn and Hilliard \cite{Cahn_Hilliard} for cases where boundary effects are neglected can be extended to problems that involve surface effects e.g. critical point wetting. However as explained in our paper \cite{Lionel} neglecting this term will not change the qualitative conclusions drawn by Cahn \cite{cahn_wetting}.

\bibliography{CommentOnCahn1977}

\bibliographystyle{plain}
\end{document}